# Design and Implementation of Electronic Infrastructure For Academic Establishment

Omar Ali Athab*        Ahmed Mohammed Saheb*

*Information and Communications Dept.
 College of Al-Khwarizmi Engineering
 Baghdad University

**Abstract**

Most establishments, including academic institutions, undergoes the lengthy process of paper-based document handling such as direct mailing, indexing and tracking. This daily task is time consuming and resource-intensive. Using a private network dedicated for such document management would benefit the establishment increasing operational efficiency. In this paper, the Information and Communication Engineering [ICE] department was used as a model to determine the requirements needed to build the Intranet network. A Packet Tracer simulator was used to build a virtual intranet architecture. Then the simulation report was examined to ensure optimum functionality. Upon establishing a stable behavior, an intranet infrastructure building commenced using the available hardware components and software. The system architecture was based on Windows 2012 R2 server to manage three separated sub-networks connected to three switches and one router. Running the intranet for one semester proved its success in providing a fast, cheap and simplified service for all department needs. The accomplished system is a step forward to achieve a full electronic department in scientific establishments.

*Keywords: intranet, computer network, network design, network simulation*

## 1. Introduction

The manual handling of documents can exhaust the establishment's resources which could be otherwise used in a more productive manner, not to mention the increased possibility for human error, leading to document loss or improper indexing. Moreover, with the numerous usage of email as a communication tool, institutions have turned on themselves to further set of problems. According to the aforementioned motivations, large establishments are gradually detecting the worth of private electronic services, the intranet, in their missions. Additionally using a local network is more secure and less susceptible to such problems (Averweg 2012) (Barraclough et al. 2009).

Essentially, the idea of Intranet is introduced to provide the requirement for communication between the workers of an establishment (Boutaba et al. 1997). Intranet is a special purpose computer network that make use of internet protocols and the available technologies to share information and communicate people through the entire connected networks of the same organization. In addition, it is intended for collaborating the activities of the work of all persons involved in that foundation (Rockley 1998).

Formerly, a number of researchers claim to have invented the idiom "intranet" that defines the growth of an internal computer network with client/server mode and based on web technology. However, around 1994, Steve Telleen (an expert in Amdahl Company) introduced the methodology of Intra Network in one of his papers. This idea, then, realized truly in Amdahl's intranet (Telleen 1996). In April 1995, the term of intranet was commercially appeared for the first time in an article, authored by Stephen Lawton (Lawton 1995). Stephen Lawton stated the pioneer international companies that have web pages and utilizing the FTP and TELNET services. Forrester Report (T. Pincince, D. Goodtree 1996) defined the term "full-service intranet" as the internal network that provides the standards-based services: e-mail, print, file, and network management. Netscape Corporation gives more details on this notion and suggests two categories of the full intranet services: network services and user services (Andreessen 1996). Network services fragmented into security, replication, directory, and management. User services involve communication and collaboration, navigation, information sharing and management, and application access. D. F. Dakhlan, et. al. (Dakhlan et al. 2015) designed and implemented an intranet structure for phasor measurement units (PMU). This intranet optimized for analysing the monitored power system stability; but with high delay. Y. Hongyin and Q. Xuelei (Yan & Qi 2011) study the implementation and key technologies of the search engine of an intranet system. They implemented a search engine for intranet depending on Apache Lucene. Experiments show that the system has a good indexing and retrieval efficiency and performance. Y. Gadallah (Gadallah et al. 2015), thought that the IoT (Internet of Thing) can be imagined as a combination of many intranets of Things (ioT). Each (ioT) belong to an association and may contain several applications. He presented a framework for the administration of existing static and mobile wireless sensor network (WSN) applications.

There are few scientific literatures on the practice of intranet implementation. In spite of the big number of researches on information systems in establishments exists, approximately all of these researches are involved with methods and structures targeted the software-based management strategies. A little amount of papers and searches methods of Information Systems management linked to implementation. (Averweg 2012)

However, one of the possible Intranet network applications is supporting course related events at academic institutions. In this context, the academic establishments need to address some issues like the information flow and the communication among the staff themselves or between instructors and the students. The current ways of communication (Martínez et al. 2006) had students complaining on the information and communication that they were given. By starting an Intranet containing information about courses such as schedules, syllabuses, exercises etc., among other things, these problem can be mitigated. It is expected that Intranet networks are gradually substituting old conventional mailing systems such as document processing systems and Bulletin Board Systems (BBS). (Kuang et al. 2013) (Lin et al. 2014)

Thus, the aim of this research is to design and implement an intranet infrastructure based on the existing technologies and using the available structure in the Information and Communication Engineering (ICE) department at Baghdad University. The implemented intranet is used to perform most of the administrative activities in the ICE dept. Consequently, by providing the computer based system, less manual labour will be needed and the speed rate of information exchange will increase. Additionally, an easy and secure communication will be achieved.

## 2. Intranet Requirements and Design

Probably the briefest definition of an intranet would be the employment of Internet tools to meet the demands of a certain group or organization (Norris et al. 1999). Therefore, the interest in this research encompasses all the academic and administrative issues for course-related events inside the ICE department. These activities can be classified into the following parts:
- Information authoring and sharing,
- Audio / Video (AV) real-time communication,
- Browsing and file/text indexing,

- E-mail,
- Different directories (of staff, courses, students, etc.)

The first step was to gather the requirements and information to find the correct approach to design the intranet network. A plan was then created showing the process and expected challenges and how it can be best tackled. Finally, the packet tracer package will be used to prepare the simulation of the intranet.

The collected requirements of the ICE dept. were:

**A) Infrastructure requirements:** the intranet service coverage should include the entire three sections of the ICE Dept. These sections are:

1. The long corridor containing the Head of the department, secretary, library, classrooms and instructors' rooms.
2. The laboratories complex (Programming and Networking laboratories).
3. The Electronics and Communications laboratories (located near but in separated building).

**B) Students Requirements:** includes the updated week course schedule, examination announcements, tutorial sheet manuals, finance and administration affairs related to students, lecture notes, text books … etc.

**C) Staff requirements:** includes the replacement of most paper-based administration works and the manual transfer of official circulars.

Based on the requirement, the research was divided into three main parts or sub-networks according to the location of the connection as shown in figure (1).

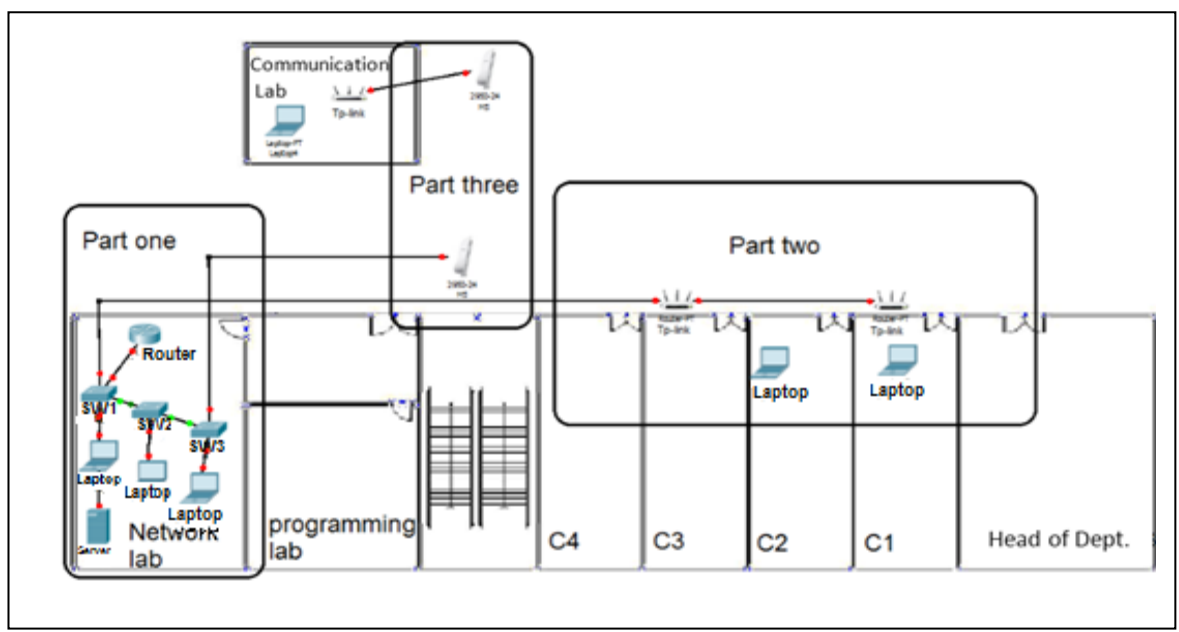

Figure (1): the general architecture of the designed Intranet

The following is an outline of the main parts and sub-networks of the work:

- The first part (Part I) is inside the network lab (it is considered the data center of the Intranet) which has a router, three switchs, laptop as a server, and several laptops as Intranet clients. these devices were connect by Ethernet cables.
- The second part (Part II) is located in pathway of the department. Two TP-Link wireless routers are used to provide coverage of the Intranet services in the pathway and classroom for students, employees, and lecture rooms and offices of the ICE department.
- The third part (Part III) is point-to-point connection to provide the intranet service to the communication lab by using two nano station transceivers.

## 3. System Simulation

The aim of intranet simulation is the preciseness investigation of all connections, sub-netting and addressing of the planned network. The simulation was also used to test the operation of some intranet services, like email and FTP web page. In addition, the simulator gives an excellent overview of all hardware and software components necessary to fulfill the specified system requirements. Finally, the simulation facilitates discovering the technical problems that may appear in the actual operation of the network.

### 3.1 Part I:

To achieve the collected intranet requirements, Part I should consist of a central server, three switches and a router. Each switch has been divided into virtual local area networks (VLANS). The purpose of using VLAN is to separate users into individual network segments for security and to provide them with their own virtual network switch. By using this technique, a lot of extra work of giving each user a separate cable for upstream or backhauling between offices were avoided. VLANs, enable us to connect a single interface on the router to **a Trunk mode port** on each switch, which allows the router to internally route between virtual VLAN interfaces, this is called logical interface. Trunk ports are also used to link between switches.

The following steps are used to set up the VLANs and to automate the process of propagating the VLAN information among the CISCO switches. The command line interface of the switch was utilized to enter the following code:

*Switch>enable*
*Switch#configure terminal*
*Switch(config)#vlan **VLAN-ID***

*Switch(config-vlan)#name **VLAN-NAME***
*Switch(config-vlan)#end*
Each VLAN must have IP address, which is given by using this code:
*Switch(config)#int vlan **VLAN-ID***
*Switch(config-if)#ip add **VLAN-IP    VLAN-SUBNET MASK***

Although VLANs operate on layer 2, passing traffic from one VLAN to another VLAN involves the use of a Layer 3 router. To setup the logical interface in the router, the following code which is written in the command line interface of the router:

*ISR(config)#int f0/0*
*ISR(config-if)#ip address  **IP    SUBNET MASK***
*ISR(config-if)#no shutdown*
*ISR(config-if)#int f0/0.**VLAN-ID***
*ISR(config-subif)#encapsulation dot1q **VLAN-ID***
*ISR(config-subif)#ip address  **IP   SUBNET MASK***

**Trunk port configuration in a switch:**
*Switch>enable*
*Switch#configure terminal*
*Switch (config)#int **NUMBER OF INTERFACE***
*Switch (config-if)#switchport mode trunk*

The next step is setting up the IP configuration for each laptop. This network has a private IP because it is an intranet network. For now, the required is only 23 hosts for each VLAN. To cope with the demand of possible future expansion of the ICE department, class C addressing was used, which provides 254 hosts. The remaining IP addresses can be used in the future to provide the service to more users. Figure 2 shows the IP configuration process.

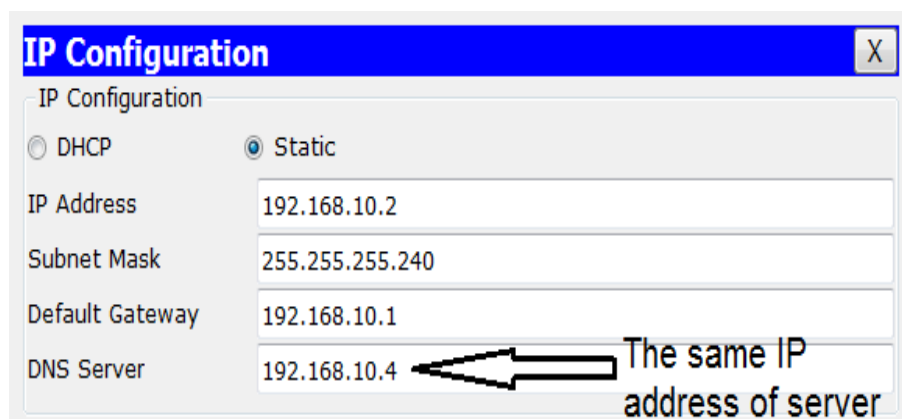

Figure (2): IP addresses given for PCs

These configurations were used for each laptop but with different IP address and different VLAN which depends on the location of each laptop. the same DNS IP address in the configuration was written for all the laptops in the network.

Part I was finished by setting VLANs, trunks for the switch and logical interface for the router. Figure 3 shows the final design of part I (data center); and each laptop was labeled with its IP address.

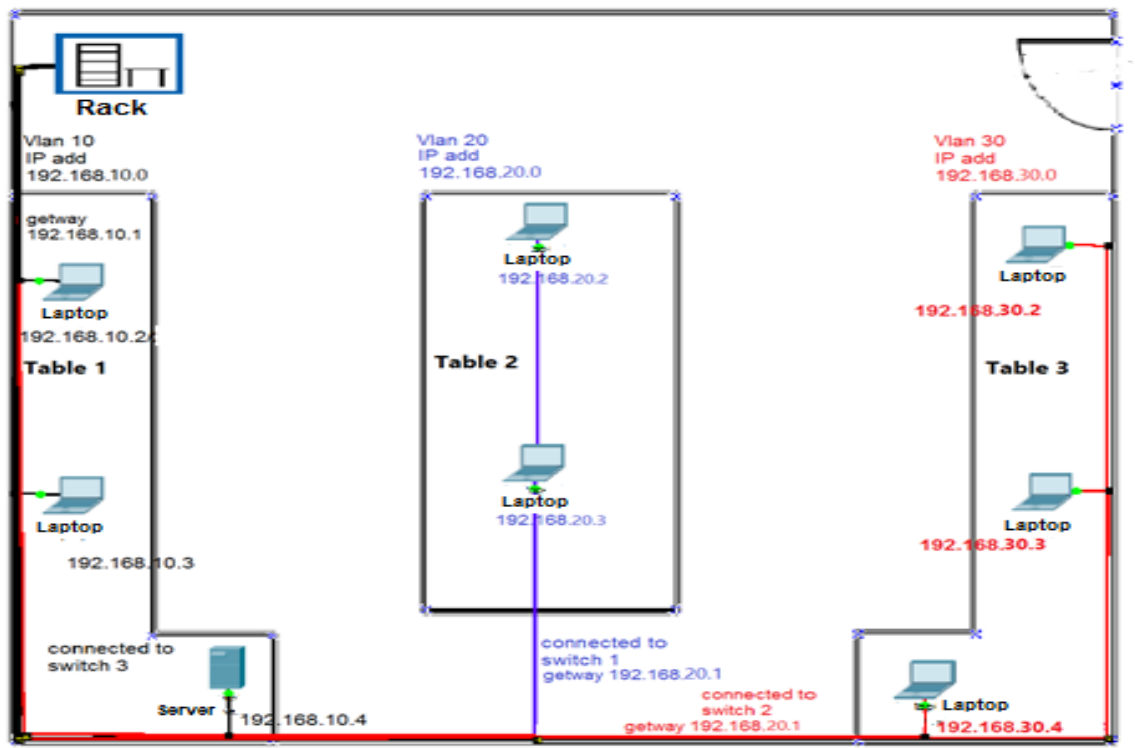

Figure (3): the Network lab (Part I) design

### 3.2 Part II :

This part provides the service to the users in the pathway and offices, a wireless router is connected to the switch VLAN10 which is located in the data center (the Network lab in part I). The router was configured as a DHCP server in order to provide all clients with an IP address. A cable is connected from VLAN10 switch port to the internet port of the wireless router. Then, another wireless router was used to extend the wireless coverage for more users in the area as an access point. The two wireless routers were conneced by using an Ethernet cable; connected from the Ethernet port of the first wireless router to the internet port of the second wireless router. The IP addresses range is 192.168.1.100 to 192.168.1.149. The DNS was configured to be 192.168.10.12. Finally, the wireless network was renamed to be (info.local) and password was set. The configuration steps for the wireless router are shown in figure 4.

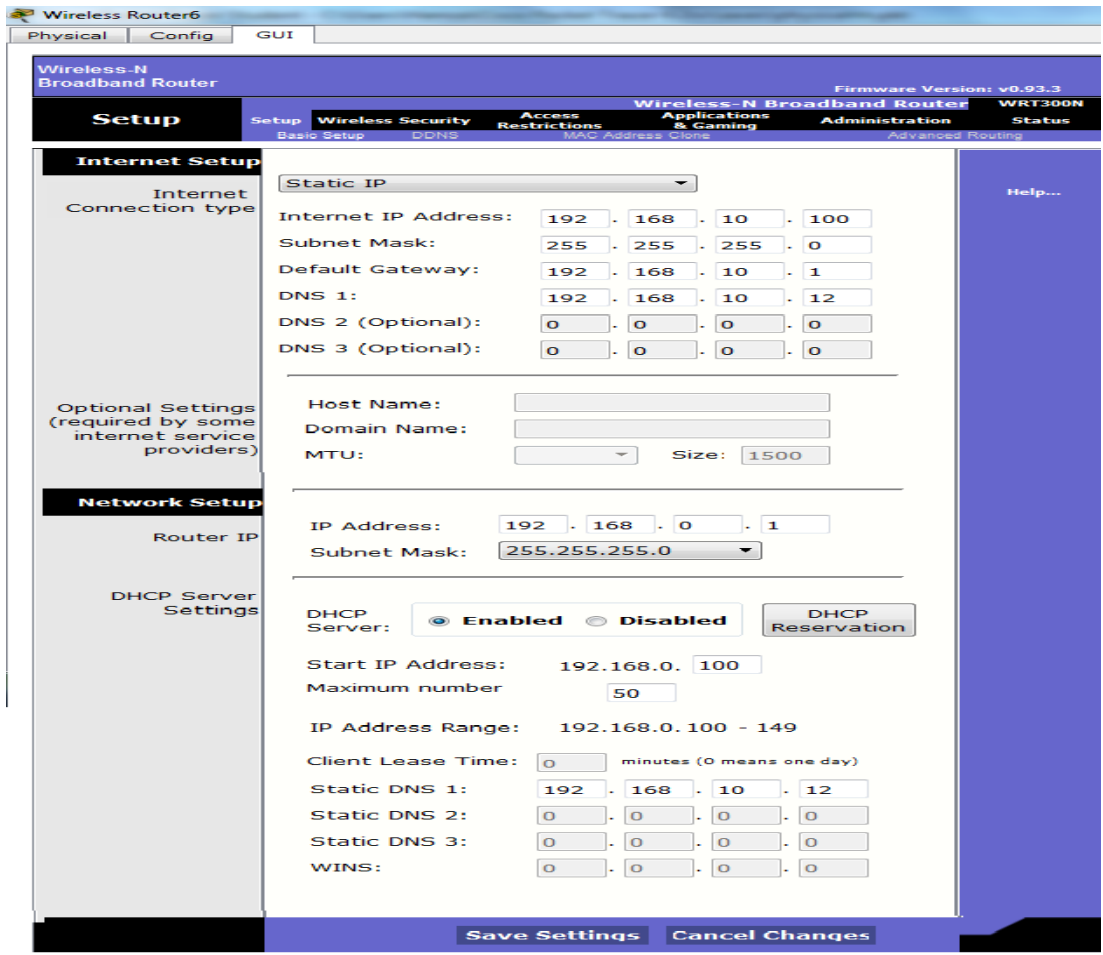

Basic setup

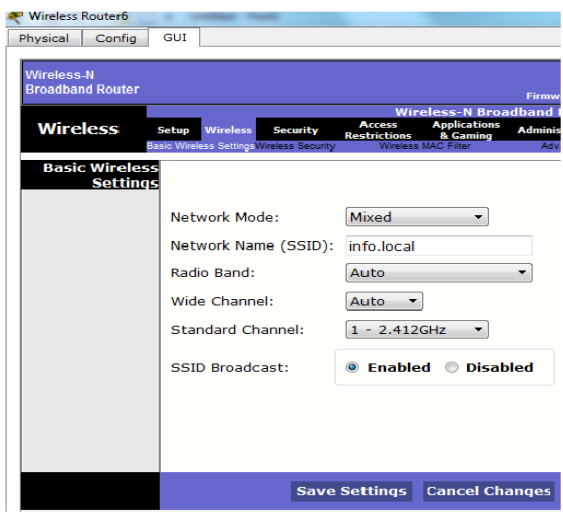

Basic wireless setting

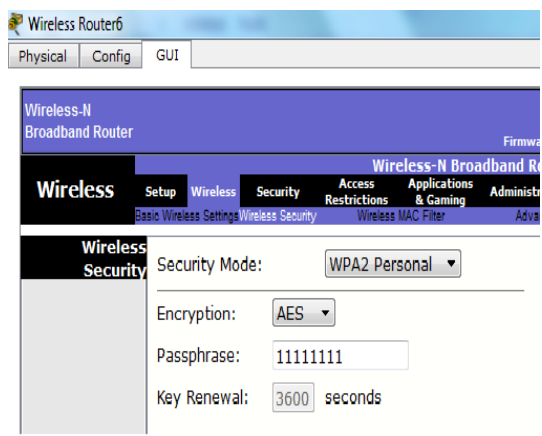

Wireless security

Figure (4) the wireless router setting

### 3.3 Part III :

This part provides the intranet service to the communication lab by using two Ubiquiti NanoStation's transceivers connected using point-to-point topology. One of the NanoStations was connected to the VLAN10 switch by using Ethernet Cable. This NanoStation was configured to be an access point. The other NanoStation, which is located at the communication lab building, is configured as a station to communicate with the access point. Then, the station was connected to a wireless router to provide a wireless covarge to the communications lab. The station is connected to the router using the LAN port in order to use the same Network ID of the VLAN10. In this wireless router, the DHCP and DNS configurations are similar to that configurations used in the first wireless router (in Part II). Figure 5 illustrates the three parts simulated using packet tracer.

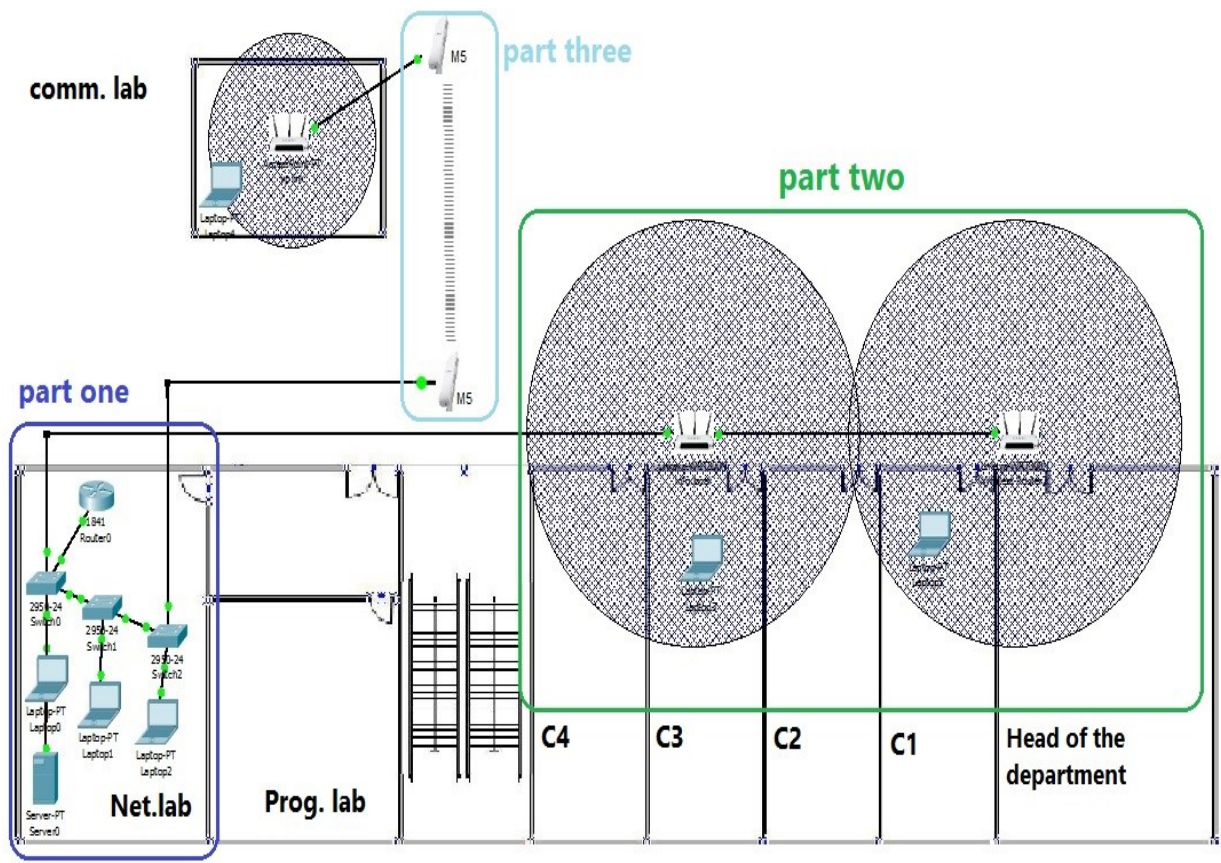

Figure (5): The final design for the simulated intranet in ICE department

## 4. System Implementation
### 4.1 System Setup
As has been mentioned in the previous section, the work was divided into three main parts. The first part is the intranet core. The second part is the lecture rooms and the offices. The third part is the communication lab. Below is a brief explanation of the implementation of each part

> A. The Implementation of the Intranet Core (Part I): The intranet core is the data center, which is located in the network lab at ICE dept. This data center consists of: one server, three switches, one router, Ethernet cables, and twenty-three laptops, as shown in figure 6. The work started by setting and installing the network. Three Cisco 2950 switches that have twenty-four ports on each one of them was connected. Each switch was divided into three VLANS (VLAN10, VLAN20, and VLAN30). Then, the server and all the laptops were connected to the three switches by using the Ethernet cable. The switches were connected to each other by a cross cable using the trunk mode port. One of the switches is connected to 2821 Cisco Router by using the sub interface.
>
> B. The implementation of the Lecture rooms and the offices (Part II): In this part, A TP-Link router was connected to the VLAN10 by using Ethernet cable. Then, a second TP-Link router was connected to the first router using Ethernet cable too.

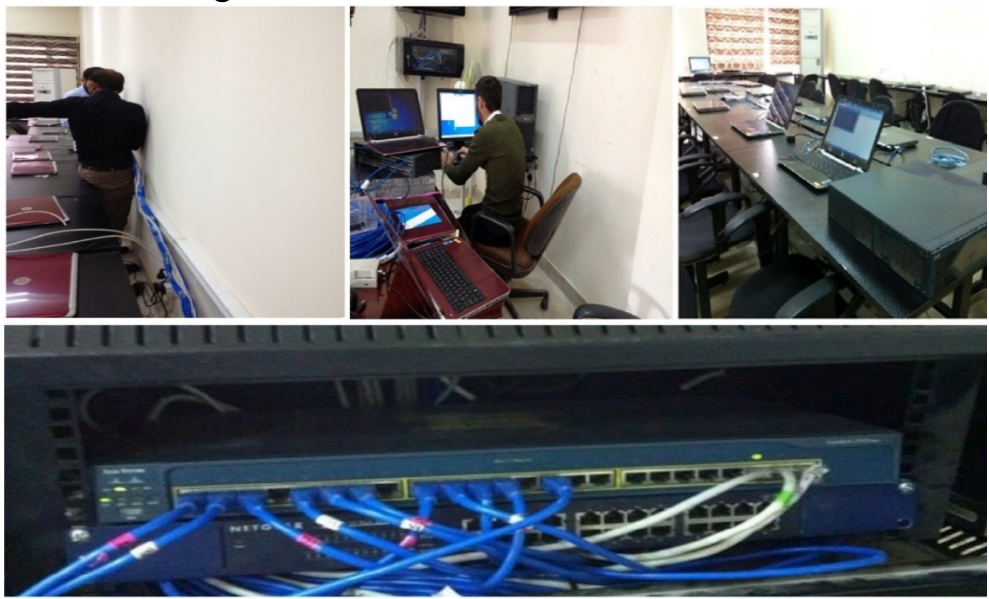

Figure (6) intranet installation of data center (Part I)

> C. The implementation of the communication lab (Part III): a NanoStation was also connected to the VLAN10 using an Ethernet cable. This NanoStation was linked wirelessly to other NanoStation located at the communication lab building using point-to-point mode. Then, the second NanoStation was joined to a wireless TP-Link router using an Ethernet cable. Figure 7 shows an outdoor view for the installed point-to-point connection.

## 4.2 Software Setting and Programming

In this subsection, the software installation and the programming that were applied to the intranet hardware; namely, to the server, Cisco router, Cisco switches, TP-Link router, NanoStation, and clients' laptops are clarified here. In the following subdivisions, the programming of each one of the hardware are listed briefly.

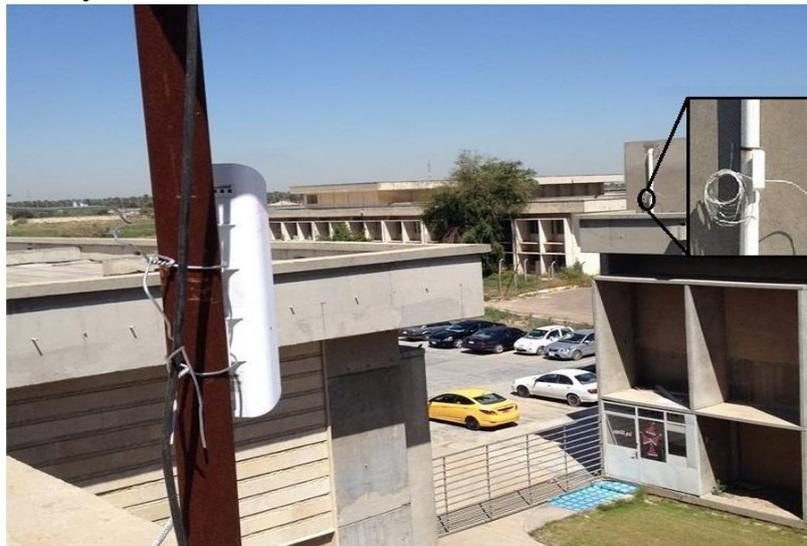

Figure 7 outdoor view of NanoStation transceivers for Part III

### A. The Cisco Router Catalyst 2821

Before the operation of the intranet, the router was configured using the flowchart shown in fig 8.

### B. The Cisco Switches Catalyst 2950

Like the router, the switch was configured using the flowchart shown in fig 9. The configurations of the second and third switches are similar to the first switch but with changing the range of IP addresses attached to interfaces for each VLAN.

### C. The TP-Link WR941N wireless router

As explained previously in section 4.1, the implementation consists of three TP-Link wireless routers; two of them are in the ICE dept. corridor (part II) and the third is located in the communications lab (part III). Each wireless router was prepared as in the following:

C.1- The default IP address replaced by a suitable one; such that it matches the sub-netting of intranet switches.

C.2- The DHCP service was enabled, and the range of IP addresses offered by each router was precisely inserted.

C.3- The IP address of the primary and secondary name servers of DNS service were set.

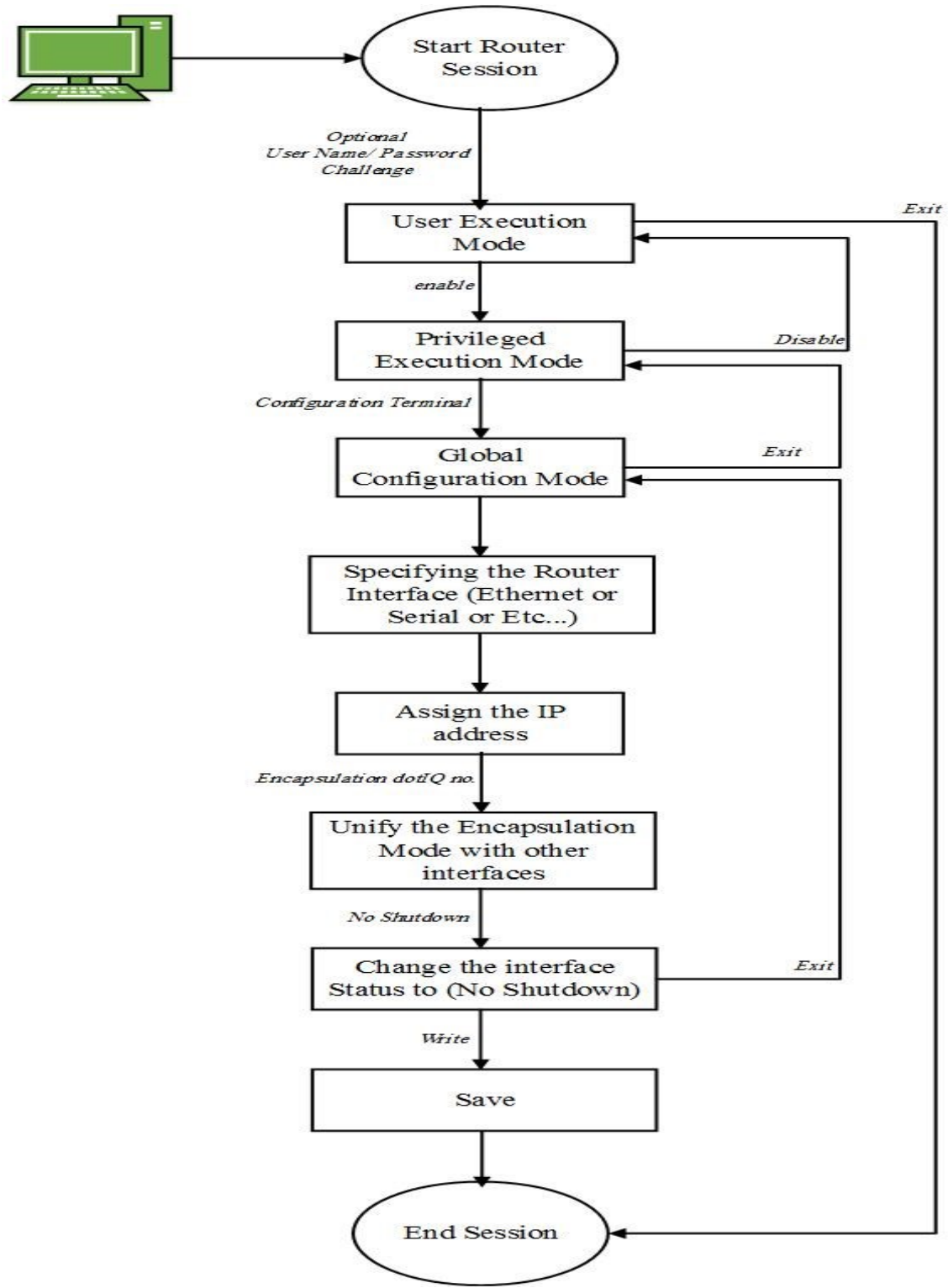

Figure 8 Router configuration flowchart

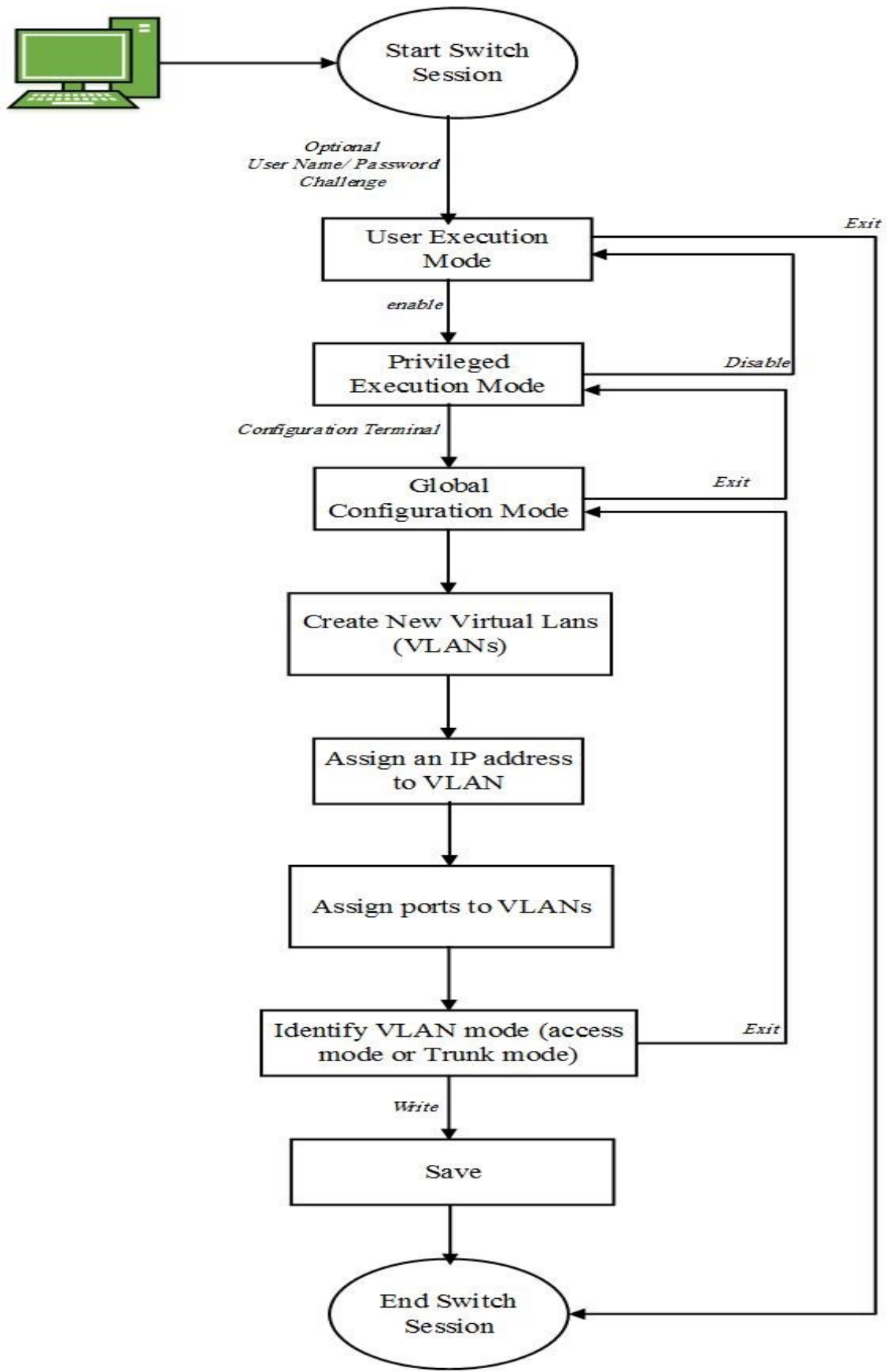

Figure 9 Switch configuration flowchart

C.4- Setting the default gateway IP address to coincide with the ICE dept. ISP gateway.
C.5- Defining the SSID (Service Set IDentifier) network name to be info.local. Due to space limitations, only some of the configuration steps of one of the three devices are shown in figure 10.

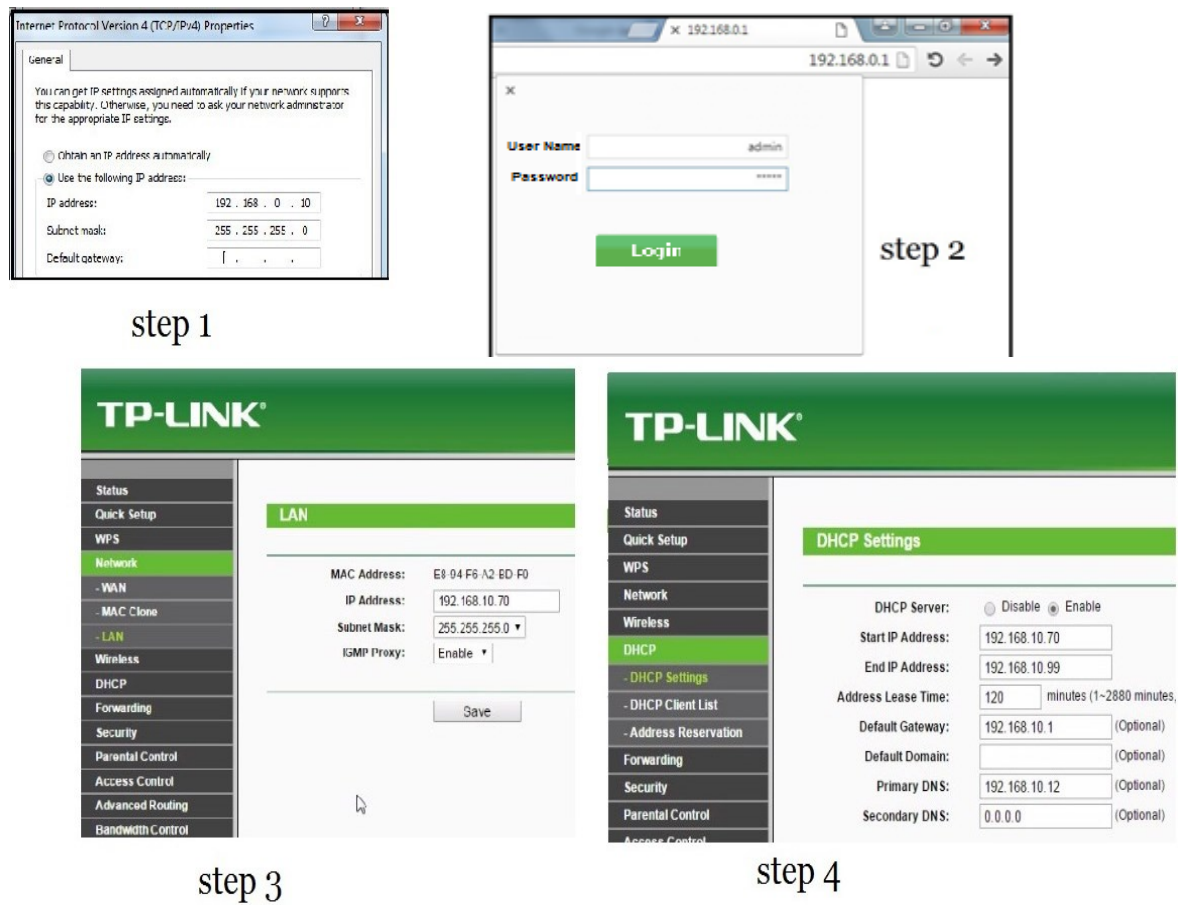

Figure (10) some of the configuration steps of TP-Link router

### D. The NanoStations M5

The first step was installing the two NanoStations at the roofs of the network lab building and the communications lab building. The mode of the first NanoStations was configured as an access point; whereas the second as a station with bridge network mode for both devices. The SSID of both NanoStations were changed to info.comm to distinguish it from indoor network (info.local). The IP addresses of both NanoStations were selected to be static IP 192.168.10.50 which matches the intranet switch sub-netting. Figure 11 summarizes the NanoStations configuration of the wireless station when linked to the access point.

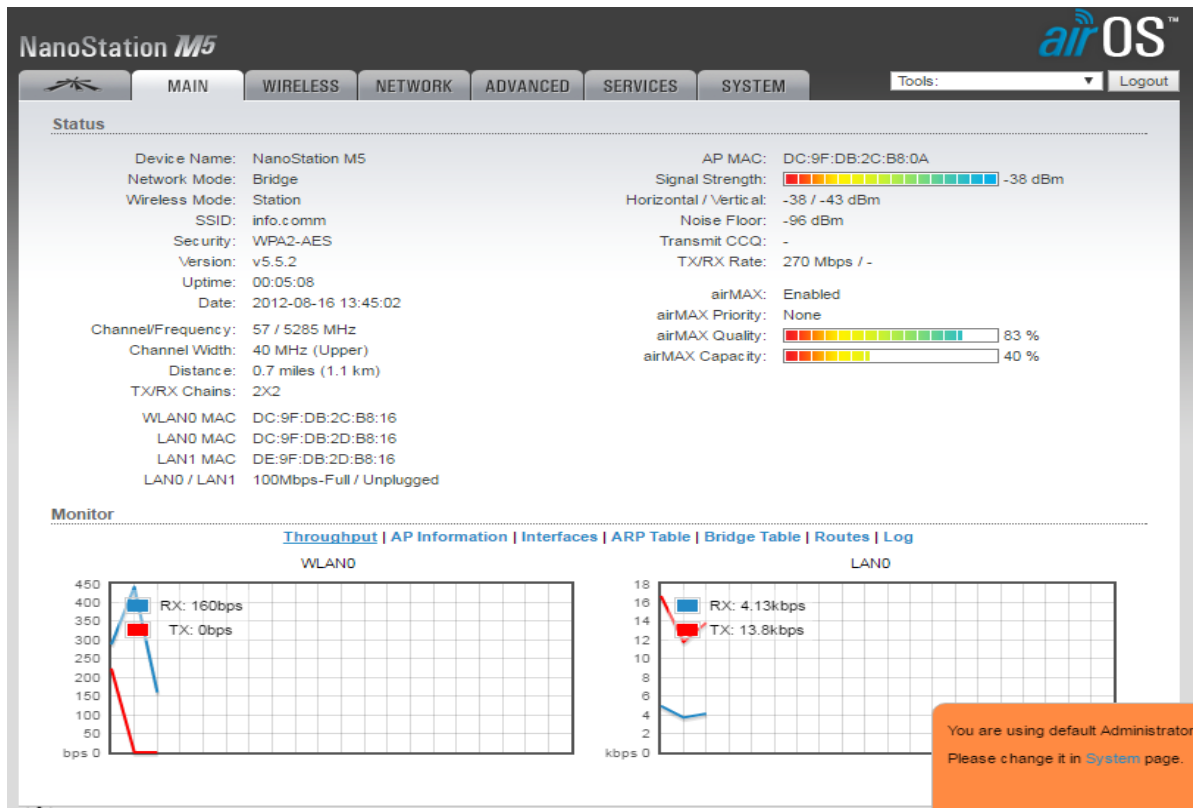

Figure (11) status of connecting the two NanoStations

### E. The Intranet Server

The server computer of the implemented intranet is located in the network lab. (Part I). Windows server 2012R2 is used as a platform. The services enabled in this network were:

**E.1** HTTP service: the HTTP service is needed for providing the web page of the intranet. This ICE dept. web site was built as part of the intranet work and it will be displayed in any client as soon as requested by its URL.

**E.2** DNS service: DNS server is enabled for the intranet web server (local.info) such that the clients can reach the ICE web site using name domain instead of IP address.

**E.3** Mail service: mail server is enabled and the Microsoft Outlook software is used as a platform. After that, the domain name of the mail was inserted to the server (mail.info.local) and then added (this mail domain and its IP address) to the DNS server so it can be resolved easily. Then, an email account in the server (user name and password) has been created for each user.

**E.4** FTP service: the FTP server is enabled by adding a name and password for the account that will be shared to the users. Only the authorized users can upload files. FTP service is used for providing lecture notes, text books, laboratory manuals …etc., for students.

**E.5** The Active Directory Domain service (ADDS) is very important to administrate the users and distribute roles for each network member. ADDS in Windows 2012 R2 is very complicated and exhaustive. However, it is used in this work to create intranet users and classify them into four groups: Students, Lectures, Employers and administrator. Each group given the suitable role through certain authorization procedure such that the requirements stated in section 2 were fulfilled. All Intranet users are given a username and password, and inserted in the ADDS. Then, the students in each studying stage rearranged into small sub-groups under the Student Group. Some of configuration steps are shown in figures 12 and 13. In this extent, the ADDS advanced tool called "Group Policy Management" is employed to manage the role for each group and sub-group, as shown in figure 14.

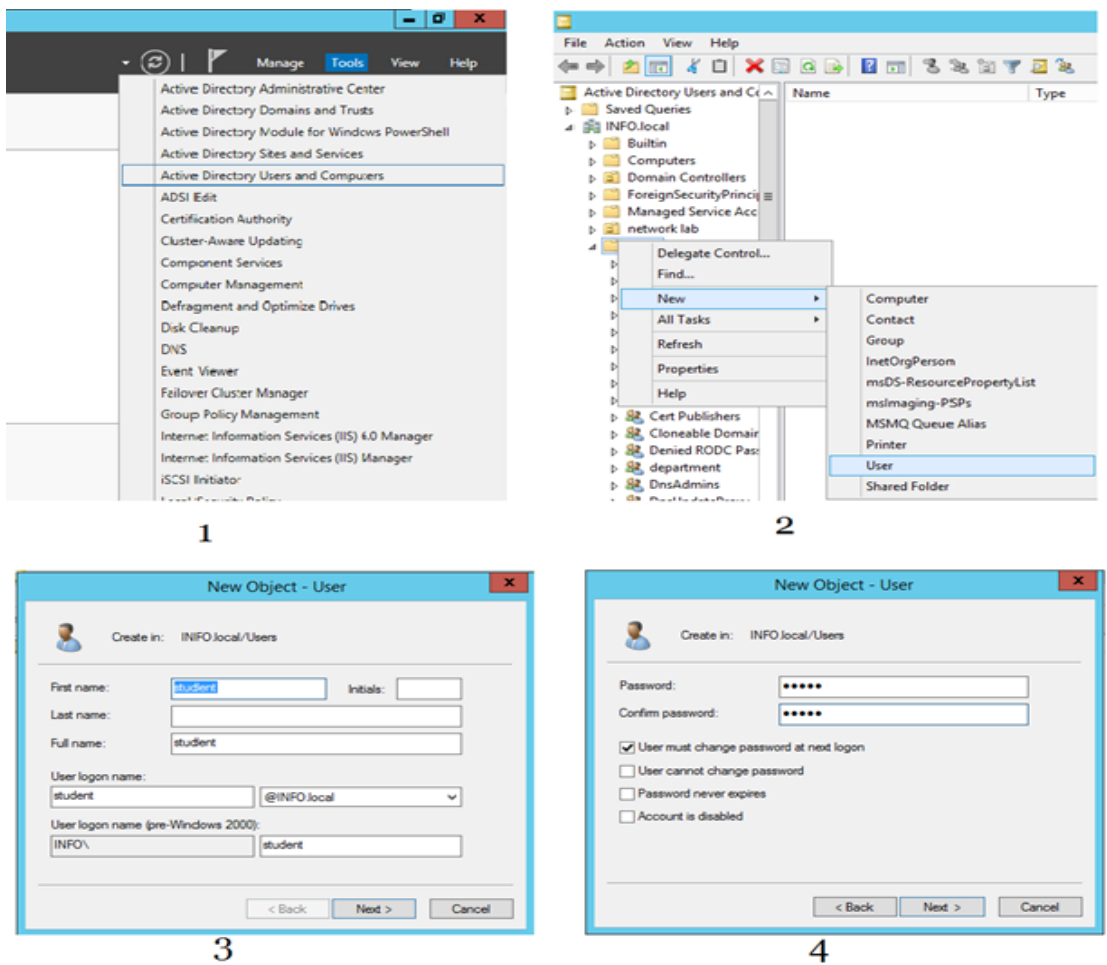

Figure (12) steps of creating users

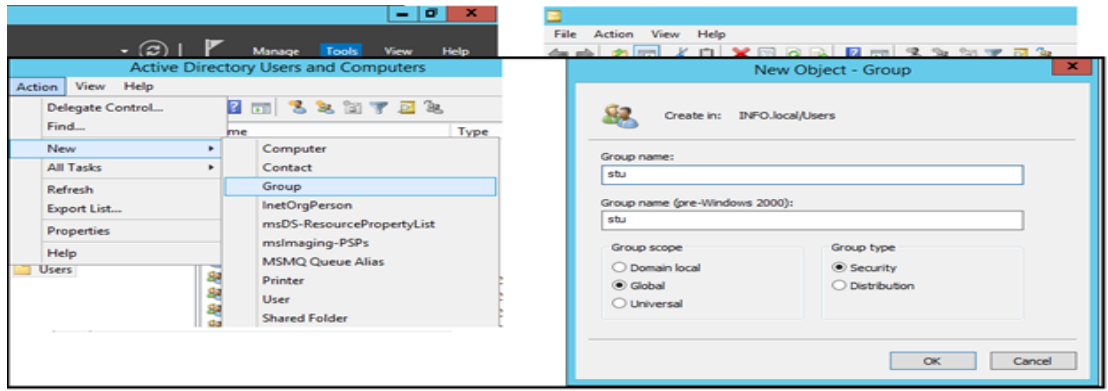
Create group

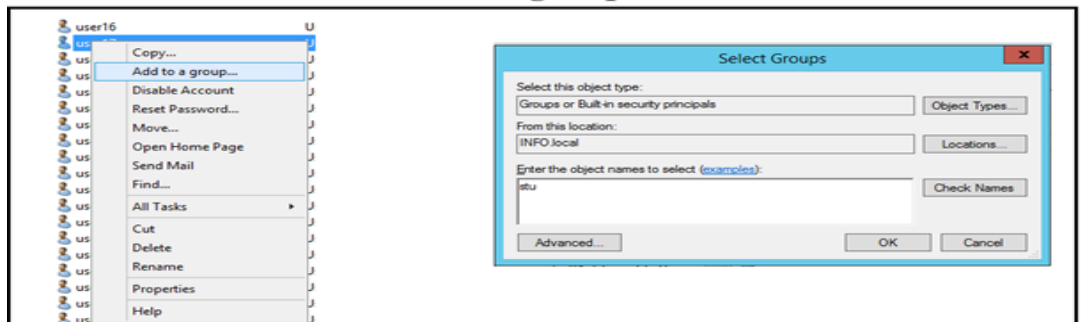
Add user to group

Figure (13) steps of creating groups

## 5. Discussion and Conclusions

The implemented intranet was operating for one semester through which the network proved its resilience, high availability and simplified service for all department members. During operation, the network had the capability of integrating devices and processes with people to ensure a highly efficient and sustainable use of resources. Using off-the-shelf equipment available at the department's stock resulted in making an affordable and simple solution. The network was capable of combining and indexing all types of information (internal and external) in one unified and accessible portal for increased ease of use.

Furthermore, the intranet proved to be a quicker information retrieval tool than the use of conventional methods. With effective search engines and its flexible management tools, users were able to quickly manipulate significant amounts of data avoiding long paper-based processes of indexing and listing. Sensitive information and circulars can be circulated instantly and securely, and a prompt feedback can be requested speeding up the process. Consequently, reaching highly optimal use of resources, which is one step closer to achieve the department's vision of creating a fully-fledged electronic department [E-Department]. However, the implemented scheme in this research is a preliminary model; hence, there is a lot of opportunities for enhancement to be able to meet any future requirements creating a future-proof and reliable intranet.

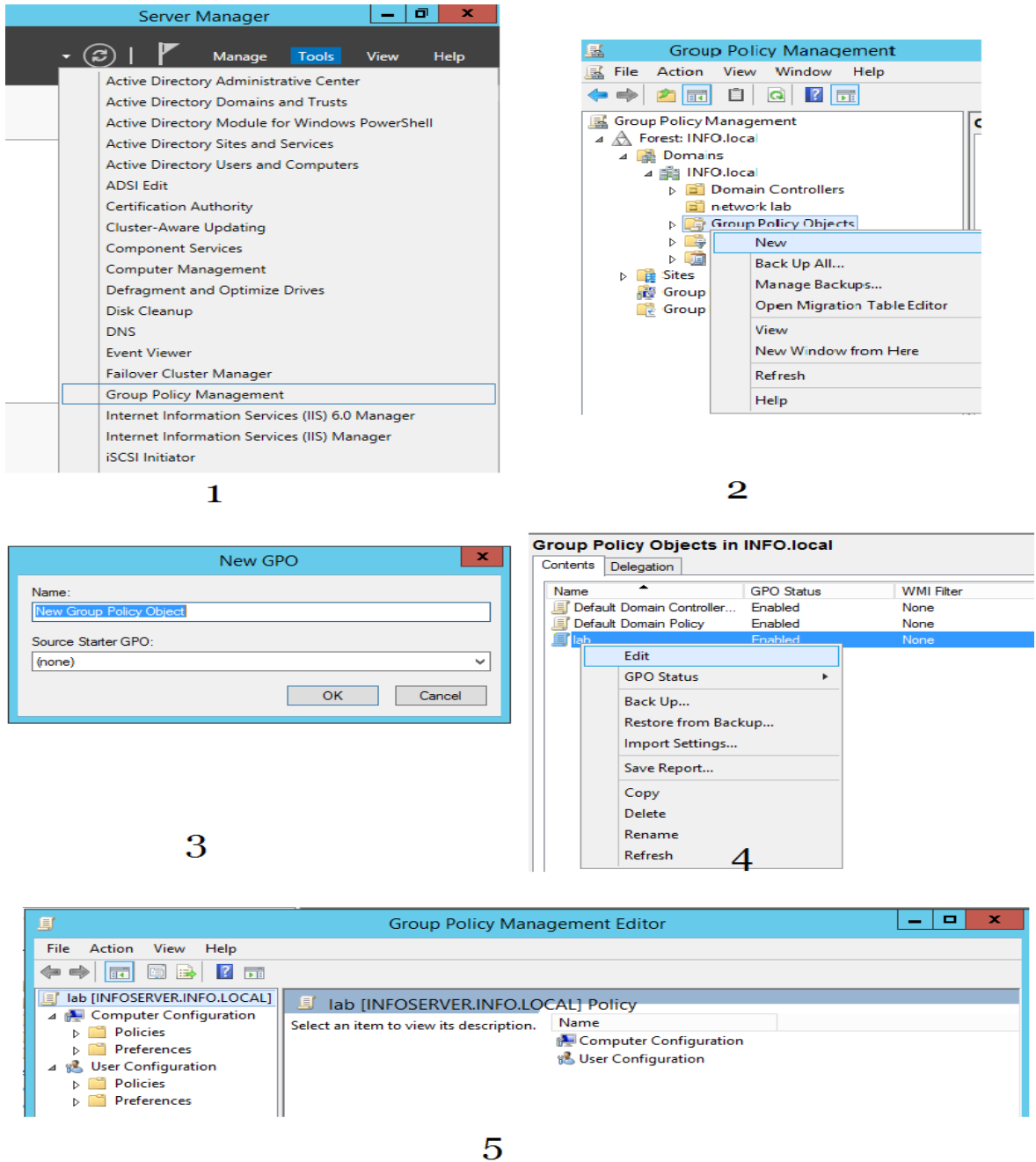

Figure (14) creating and editing group policy